\documentclass[english]{article}
\usepackage{graphicx,color}
\usepackage{amssymb}
\def\no{\noindent}
\def\bc{\begin{center}}
\def\ec{\end{center}}

\def\beq{\begin{equation}}
\def\eeq{\end{equation}}

\def\d{\downarrow}
\def\u{\uparrow}

\def\br{{\bf r}}

\def\bk{{\bf k}}

\def\bc{{\bf c}}

\begin{document}

\title{
Analytic bulk-edge connection in circular-symmetric models
}

\author{K. Ziegler\\
Institut f\"ur Physik, Universit\"at Augsburg\\
D-86135 Augsburg, Germany\\
email: klaus.ziegler@physik.uni-augsburg.de
}

\maketitle

Abstract:

We propose a systematic analysis of the eigenfunctions of two-band systems
in two dimensions with a circular edge. Our approach is based on an analytic 
continuation of the wavenumber, which yields a mapping from the bulk modes to 
the edge modes. Phase relations of the eigenfunctions are described by their 
mapping onto a three-dimensional field of unit vectors. This mapping is studied 
in detail for a two-band Laplacian model and a Dirac model. The direction of the unit 
vector identifies the phase relation of the eigenfunctions and enables us to distinguish
between the upper band, the lower band and the edge spectrum. Bulk and edge modes
are spectrally separated, which results in two transitions from delocalized 
bulk modes to localized edge modes. These transitions are accompanied by transitions
of the phase relations.
Our analytic approach is compared with the topological bulk-edge correspondence, which is based
on the Chern number of the bulk.

\tableofcontents

\section{Introduction}

Recent studies on complex photonic and phononic systems have opened up new perspectives 
for analyzing wavefunctions, which are solutions of the Schrödinger or Dirac equations
of single quantum particles~\cite{PhysRevLett.100.013904,Lu2014,RevModPhys.91.015006,Ni2023}. 
Besides the spectrum, wavefunctions carry 
crucial information about both classical and quantum systems, particularly regarding 
their topological properties, which influence optical and transport behaviors. 
In contrast to electronic systems, wavefunctions can be directly observed in classical 
systems, and their coherence is often easier to manipulate 
than in electronic systems. For example, the intensity of an electromagnetic or 
sound wave, represented by the magnitude of the wavefunction, can be measured locally at 
position $\br$. 
Moreover, the structure of the two-component wavefunction of a 
two-band Hamiltonian offers even deeper insights, since the relative phase between the 
two components plays a central role in determining the system's properties.

In a system with edges we can distinguish bulk and edge modes. Both are eigenfunctions
of the Hamiltonian but for different eigenvalues and with qualitatively different properties: 
the bulk modes are wave-like functions, extended over the entire systems, while the edge modes decay
exponentially away from the edge. 
Edge modes play a crucial role in the quantum Hall effect~\cite{halperin82,gruber06}. They are quite robust but
at the same time they are also sensitive to the boundary conditions.
Several types of boundary conditions and their corresponding edge modes were studied in electromagnetic
systems~\cite{Ziegler:18}. 

The goal of this work is to analyze the connection
between those two types of modes, which is related to the bulk-edge connection. 
A similar idea to describe the bulk-edge correspondence was pursued for
topological materials, based on the connection of topological invariants of the bulk (Chern numbers)
and the number of edge modes. In particular, this has been discussed intensively in the context of the 
quantum Hall effect and the 2D Dirac equation, pioneered by the work of Hatsugai~\cite{
hatsugai93}, and followed by a broader application of this idea in 
Refs.~\cite{doi:10.1142/S0129055X20300034,Graf2013,doi:10.1126/science.aan8819,Graf2018,PhysRevB.99.035153,
10.1093/ptep/ptaa140,Li2023,bal23,PhysRevResearch.6.033161,Wong2024,Isobe_2024,
2403.04465,rossi24,2410.13940}.

In the following we will develop a more direct approach for the bulk-edge connection, using an analytic continuation
without relying on topological invariants.
For this purpose we consider the equation $H\Psi_E=E\Psi_E$ with $\Psi_E=(\psi_{E,1},\psi_{E,2})$
for some models with a two-band Hamiltonian $H$, which is assumed to act on a two-dimensional 
space and has the general structure 
\beq
\label{2band_ham00}
H=
h^{}_1\sigma^{}_1 + h^{}_2\sigma^{}_2 + h^{}_3\sigma^{}_3
\equiv {\vec h}\cdot{\vec\sigma}
\eeq
in terms of Pauli matrices $\sigma_j$ with ${\vec\sigma}=(\sigma_1,\sigma_2,\sigma_3)^T$.
We focus on a circular symmetry, which is easy to realize experimentally.

The analytic continuation of wavefunctions is a powerful method, which has been applied to 
many physical problems. A typical example is the scattering theory, where bound states are 
found as poles of the scattering matrix or the Green's function. In this paper we will solve
the eigenvalue problem $H\Psi_E=E\Psi_E$  for the bulk modes
and construct the corresponding edge modes by an analytic continuation, extending an
idea employed to a 2D superconductor~\cite{ziegler2024edgemodeschiralelectron}.

\section{Two-band models}

Inspired by the electronic properties of graphene~\cite{Novoselov2005} as well as by wave properties in
photonic and phononic systems on a honeycomb structure, the 2D Dirac Hamiltonian
with ${\vec h}=(i\partial_x,i\partial_y,m)$ with Dirac mass $m$ in the Hamiltonian of 
Eq. (\ref{2band_ham00}) has been extensively studied in recent 
years~\cite{PhysRevLett.100.013904,Lu2014,RevModPhys.91.015006,Ni2023,Cheng2016-qk}.
The eigenfunction of a translational-invariant two-band Hamiltonian reads
\beq
\Psi_\bk(\br)=\pmatrix{
\psi_1 \cr
\psi_2 \cr
}e^{i\bk\cdot\br}
,
\eeq
where $\br=(x,y)$ is the position in space and $\bk$ is the wavevector.
Moreover, in a circular symmetric system the position is parametrized by the
radius $r$ and the polar angle $\alpha$, and the two components of the wavevector 
are parametrized by the wavenumber $k$
and the angular momentum number $n=0,\pm 1,\ldots$ such that the eigenfunction reads
\beq
\Psi_{k,n}(r,\alpha)=\pmatrix{
\psi_1 \cr
\psi_2 \cr
}\frac{e^{ikr+in\alpha}}{\sqrt{kr}}
.
\eeq

\subsection{Topological properties}

\begin{figure}[t]
\begin{center}
\includegraphics[width=0.3\linewidth]{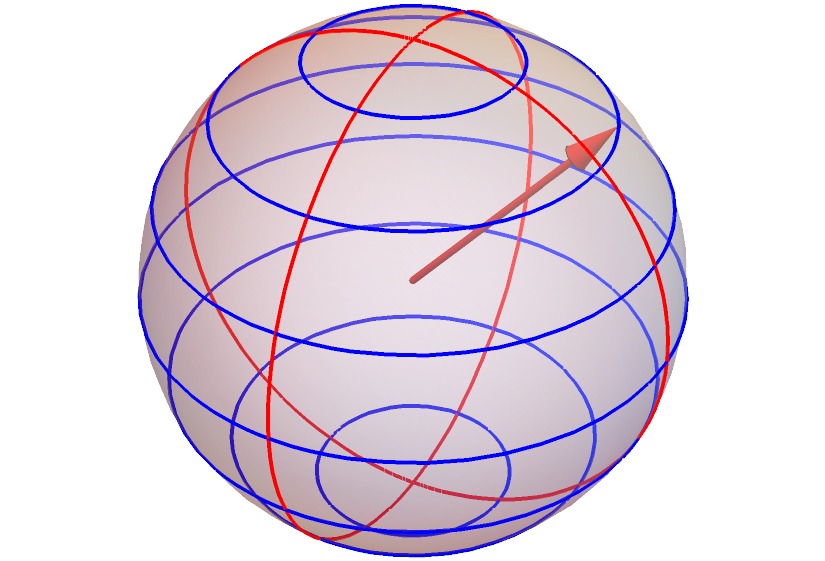}
\caption{
Sphere of the $S^2$ field ${\vec s}$.
}
\label{fig:sphere}
\end{center}
\end{figure}

Topological properties of the two-component wavefunction $\Psi(\br)=(\psi_1,\psi_2)^T$,
such as its chirality, can be analyzed through topological invariants, such as winding numbers or
Chern numbers. Another interesting quantity in this context is the Hermitian tensor field $\psi_i^*(\br)\psi_j(\br)$, 
which is gauge-invariant in the sense that phase factors of the wavefunctions cancel each other
in the product and only phase differences survive. 
Using $\psi_j=|\psi_j|e^{i\varphi_j}$, we define a real three-dimensional vector ${\vec s}$
from the Hermitian tensor $\psi_i^*(\br)\psi_j(\br)$ as
\beq
\label{s3-field}
s_1=\frac{2|\psi_1||\psi_2|}{|\psi_1|^2+|\psi_2|^2}\cos(\varphi_2-\varphi_1)
\ ,\ \
s_2=\frac{2|\psi_1||\psi_2|}{|\psi_1|^2+|\psi_2|^2}\sin(\varphi_2-\varphi_1)
\ ,\ \ 
s_3=\frac{|\psi_1|^2-|\psi_2|^2}{|\psi_1|^2+|\psi_2|^2}
,
\eeq
which characterizes the eigenfunctions according to the magnitudes of their vector components and 
their phase differences. 
The vector components $s_1, s_2$ provide a winding number of the wavefunctions
through their phase dependence. Therefore, this gauge-invariant field is reminiscent of the 
Berry connection. Direct inspection reveals that ${\vec s}=(s_1,s_2,s_3)^T$
is a three-dimensional unit vector due to $s_1^2+s_2^2+s_3^2=1$. Thus, 
the trajectory of ${\vec s}(\br)$ is a horizontal circle on the unit sphere 
when we vary the phase difference $\varphi_2-\varphi_1$ from 0 to $2\pi$,
as visualized in Fig. \ref{fig:sphere}.
The mapping $(\psi_1(\br),\psi_2(\br))^T\to{\vec s}(\br)$, which reflects a mapping of the two-dimensional
plane to the unit sphere $S^2$, enables us to identify the 
wavefunction as a structure on a compact manifold. An expansion of ${\vec s}$ in terms of Pauli matrices
yields
\beq
{\vec s}=\frac{
\Psi\cdot{\vec\sigma}\Psi^{}}{\Psi\cdot\Psi}
.
\eeq
In general, the mapping ${\vec h}\to{\vec s}$ is central for a two-band Hamiltonian. 
It will be shown subsequently that ${\vec s}(\br)$ characterizes the properties
of the eigenfunctions through its trajectories when we vary the position $\br$.
Besides ${\vec s}$, the local intensity or signal strength $I=\Psi\cdot\Psi=|\psi_1|^2+|\psi_2|^2$ 
is another relevant quantity to characterize the eigenfunctions of $H$,
which yields the spatial distribution of the signal strength.
The distribution of bulk modes is quite different from the distribution of edge modes,
since for the latter it is concentrated only at the edge(s).
In classical systems the spatial integral of 
$I$ is the energy stored in the sample, while for quantum systems it is 1.

Topological properties, such as the chirality, can be identified, for instance, 
with the edge vorticity (EV) of a circular edge, which is defined as 
\beq
\label{edge_vort}
{\cal V}=\int_{\cal F}[\nabla\times{\vec s}]\cdot d^2\br
,
\eeq
where ${\cal F}$ is the two-dimensional area of a circular hole or a disk,
whose circular edge $\partial{\cal F}$ carries an edge mode. 
$d^2\br$ is the oriented differential element ${\bf n}d^2r$.
Although this is not a topological invariant, its sign characterizes topological
properties, similar to the Chern number of the band. The EV reveals local properties with respect to the edge, 
while the Chern number is a global property, since it is an integral over the entire Brillouin zone.

The phase difference in Eq. (\ref{s3-field}) depends on the location $(r,\alpha)$, and
$\nabla\times{\vec s}$ can identify a vortex. A simple example is the eigenfunction of the
translational-invariant 2D Dirac Hamiltonian:
\beq
\Psi_{\bk;\pm}(\br)
=\pmatrix{
1 \cr
\rho_{k;\pm}e^{i\gamma} \cr
}e^{i\bk\cdot\br}
\ , \ \rho_{k;\pm}=-\frac{m\pm \sqrt{m^2+k^2}}{k}
\eeq
for the eigenvalues $\pm\sqrt{m^2+k^2}$.
This gives immediately a spatial uniform ${\vec s}_\bk=2\rho_{k;\pm}(\cos\gamma,\sin\gamma,0)^T$ 
with $\gamma={\rm arg}(k_x+ik_y)$, while the divergence with respect to $\bk$
is proportional to $\rho_{k;\pm}$. This indicates a source (sink) of ${\vec s}_\bk$
at the center for the lower (upper) band for $m>0$. On the other hand, the EV vanishes due to ${\cal V}=0$. 
$\rho_{k;\pm}$ switches its sign with $m\to -m$. In the following we will assume that $m>0$.

\section{Topological bulk-edge correspondence and analytic bulk-edge connection}

Almost all experimental samples have at least one edge, where a typical example is a disk
with a circular edge. The contribution of the edge to the properties of the
sample is exponentially small and is usually ignored in the macroscopic
description. This changed after the discovery of the quantum Hall effect
with a robust Hall conductivity~\cite{klitzing}, where the latter was attributed to edge 
modes~\cite{halperin82}. 
It soon turned out that it is crucial to understand how the edge modes are related
to the bulk modes, since the existence of edge modes inside the spectral gap of the
bulk modes plays a crucial role to explain the transport properties of the quantum 
Hall effect~\cite{hatsugai93}.
A constructive approach to the bulk-edge correspondence can be based on the projection 
in $\bk$ space perpendicular to a straight edge~\cite{doi:10.1142/S0129055X20300034,PhysRevB.83.125109}.
The corresponding set of one-dimensional solutions, parametrized by the wavevector along
the edge, provides the edge modes and their energies. This approach is rather
involved though. In particular, it requires the solution of the one-dimensional equations for the edge. 
Therefore, we suggest here a different approach, based on an analytic continuation
of the wavevector, which we applied previously to the Bogoliubov de Gennes equation for
superconducting double layers~\cite{ziegler2024edgemodeschiralelectron}. 
To distinguish it from the  topological bulk-edge correspondence (TBEC), it will be called subsequently 
analytic bulk-edge connection (ABEC). An advantage
of this approach is that it is sufficient to solve only the bulk equation.  
We will argue in the following that edge modes appear inside a spectral gap in systems without
specific reference to topology of the eigenfunctions (i.e. it is applicable also to
eigenfunctions with vanishing EV). 
It will be shown that their origin is purely geometric, typically due to an edge.
They are specific for a given Hamiltonian $H$ but can be determined within a systematic
approach, which is based on an analytic continuation of the bulk modes.

Starting point is the spectrum $E(k)=(m^2+k^2)^{1/2}$ of the 2D Dirac Hamiltonian 
with real wave number $k$ and a positive Dirac mass $m$. This example is considered
for its simplicity but the concept can be directly generalized to other Hamiltonians
with the bulk spectrum $E(\bk)=(m^2+h(\bk))^{1/2}$ with an analytic function $h$. 
The square root provides two bands $\pm\sqrt{m^2+k^2}$ for the 2D Dirac case,
which are separated by a gap $2m$. Formally, this spectrum can be extended 
by continuing the real wave number $k$ to the complex plane. This creates a connected Riemann
surface (cf. Fig. \ref{fig:riemann}a,c). Therefore, the gap is bridged by modes with complex $k$.
However, a complex eigenvalue $E$ is not physical, such that we must restrict the analytic
continuation to real values of $E$. This is the case for real $k$ and for purely imaginary 
values $k=ic$ with $-m\le c\le m$, where the latter gives $E=\pm\sqrt{m^2-c^2}$. This 
means that the spectrum separates in a bulk spectrum with $E\ge m$ and $E\le -m$ 
and an edge spectrum with $-m<E<m$.

For the analytic continuation of the eigenfunctions we
assume that the bulk mode $\Psi(x,y)$ is either a plane wave $e^{ik_1x+ik_2y}$ for a 
translational-invariant system or a circular wave $e^{ikr+in\alpha}$ with polar 
coordinates $(r,\alpha)$ for a circular-invariant system.
Then we obtain from the analytic continuations $k_j\to ic_j$ and $k\to ic$
either $e^{-c_1x-c_2y}$ or $e^{-cr+in\alpha}$, respectively.
These modes decay exponentially for $c_1x>0$, $c_2y>0$, representing a so-called corner mode, or $cr>0$,
representing a circular edge mode. The boundary conditions play obviously a crucial role here,
since the edge modes should decay exponentially from the edge. This selects whether $c_j$, $c$ are
positive or negative. For a large disk with a central hole, for example, this means that $c>0$
at the edge of the hole.

We note that the analytic continuation does not affect the Hamiltonian, only the spectrum and the
eigenfunctions. The analytic continuation can be reversed such that we get the
bulk modes from the edge modes. This can be useful when we observe or manipulate the 
edge modes and want to determine the corresponding bulk modes. 

\section{Special examples}
\label{sect:examples}

\subsection{Single-band Laplacian model}
\label{sect:single_l}

We consider the eigenvalue problem $H\Psi_E=E\Psi_E$ with the Hamilton operator $H$, acting
on a 2D space, and the real energy eigenvalue $E$. This equation appears 
also in many areas of classical physics, for instance, in microwave systems, photonics and 
phononics~\cite{PhysRevLett.100.013904,Lu2014,RevModPhys.91.015006,Ni2023}.
In the following, our models are characterized by different Hamiltonians $H$.
A prototype of $H$ is a single Laplacian $H=-\Delta$ on a disk with a non-negative bulk spectrum.
For a the circular geometry $\Delta$ is parametrized by polar coordinates $(r,\alpha)$ as
\beq
\label{laplacian}
\Delta=\frac{\partial^2}{\partial r^2}+\frac{1}{r}\frac{\partial}{\partial r}
+\frac{1}{r^2}\frac{\partial^2}{\partial\alpha^2}
.
\eeq
The Bessel functions $J_n(kr)$ and $Y_n(kr)$, multiplied by a phase factor $e^{in\alpha}$, 
are eigenfunctions of this Laplacian with eigenvalues $-k^2$~\cite{abramowitz+stegun}. Two linearly independent 
solutions are given by the linear combinations 
\beq
\label{wf0}
\phi^\pm_{k,n}(r,\alpha):=
J_n(kr)\pm iY_n(kr)
=c_n\frac{e^{\pm ikr}}{\sqrt{kr}}+O\left(\frac{1}{kr}\right)
\ \ (n=0,\pm 1,\ldots)
\eeq
with $c_n=e^{\mp i(\pi n+\pi/2)/2}\sqrt{2/\pi}$. The eigenvalues $k^2$ of $-\Delta$ are
independent of $n$ due to the rotational invariance of the Laplacian.

The analytic continuation $k\to ic$ with a real $c$ yields negative eigenvalues $k^2\to-c^2$ and
the modified Bessel functions as (cf. App. \ref{app:bessel})
\beq
J_n(kr)\to J_n(icr)=e^{i\pi n/2}I_n(cr)
\ ,\ \
J_n(kr)+iY_n(kr)
\to H_n^{(1)}(icr) 
=-\frac{2i}{\pi}e^{-i\pi n/2}K_n(cr)
,
\eeq
where the modified Bessel function $I_n(cr)$ ($K_n(cr)$) increases (decreases) exponentially for 
$cr>0$ and $K_0(cr)$ diverges for $cr\to0$. A proper linear combination
gives a unique solution that satisfies the boundary conditions at the edge.
For a disk this is matched by $I_n(cr)$ with a finite value at $r=0$ 
and for a hole on a large disk it is matched by $K_n(cr)$.
The exponential decay rate of the edge modes is given by $1/|c|=1/\sqrt{|E|}$, indicating a 
radial shrinking of the edge modes with decreasing energy $E=-c^2$. 

The requirement of the analytic continuation is that (i) the 
eigenvalues remain real and (ii) edge modes decay exponentially from an edge 
(i.e., they are evanescent modes). Thus, they depend strongly on the boundary
conditions.
We conclude that the spectrum of $-\Delta$ consists of two branches, which are parametrized by
the complex wavenumber $k$. 
One branch is for the bulk modes with $E\ge 0$ and the other is
for the edge modes with $E<0$. 
Although both spectra are real, the bulk spectrum is parametrized by a real wavenumber, while
the edge spectrum is parametrized by an imaginary wavenumber.

\subsection{Two-band Laplacian model}
\label{sect:2_l}

The 2D Laplacian $\Delta$ can be used to construct the Hamiltonian in Eq. (\ref{2band_ham00})
with $h_1=-\Delta$, $h_2=0$ and $h_3=m$. This yields the real symmetric matrix
\beq
H_{\rm 2\Delta}=\pmatrix{
m & -\Delta \cr
-\Delta & -m \cr
}
.
\eeq
Since the three-component vector
${\vec h}=(-\Delta,0,m)$ is only a one-dimensional line
in 2D, the bands have vanishing Chern numbers.

\begin{figure}[t]
\begin{center}
\includegraphics[width=0.7\linewidth]{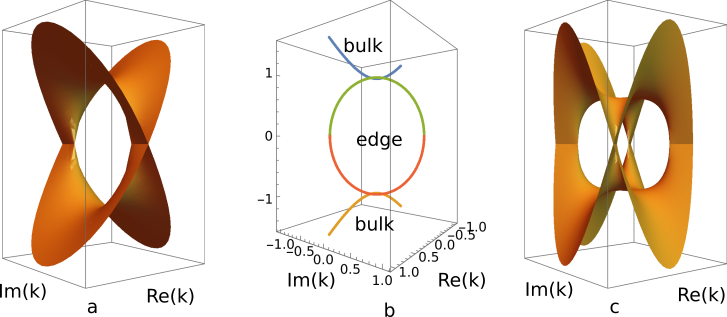}
\caption{
The real part of $E(k)$ for a complex wavenumber $k$.
a) Riemann surface of the real part of $E(k)=(m^2+k^2)^{1/2}$ and 
b) the related bulk and edge spectra for a real $E(k)$.  
c) Riemann surface of the real part of $E(k)=(m^2+k^4)^{1/2}$.
}
\label{fig:riemann}
\end{center}
\end{figure}

Now we consider $H_{\rm 2\Delta}$ on a space with a circular geometry (e.g., a disk). Using the Laplacian 
in Eq. (\ref{laplacian}) and its eigenfunctions in Eq. (\ref{wf0}), the ansatz
\beq
\label{ansatz2}
\Psi^\pm_{k,n}(r,\alpha)=\pmatrix{
a_1 \cr
a_2 \cr
}\phi^\pm_{k,n}(r,\alpha)e^{in\alpha}
\eeq
with the wavenumber $k$ yields
\beq
\label{ham_2}
H_{\rm 2\Delta}\Psi^{\pm}_{k,n}(r,\alpha)
=\pmatrix{
m & k^2 \cr
k^2 & -m \cr
}\pmatrix{
a_1 \cr
a_2 \cr
}\phi^\pm_{k,n}(r,\alpha)e^{in\alpha}
\eeq
with the eigenvalues $E_k=\pm\sqrt{m^2+k^4}$. The components of the eigenvectors $(a_1,a_2)^T$ read
\beq
\label{2b_laplacian}
a_1=k^2
\ , \ \
a_2=m\pm\sqrt{m^2+k^4}
\eeq
with an additional normalization that depends on the specific physical system.
The $k^4$ dependence of the eigenvalues suggests the analytic continuation 
$k\to c=e^{\pm i\pi/4}c$ with $-\sqrt{m}\le c\le \sqrt{m}$,
such that $E_k=\pm\sqrt{m^2+k^4}\to \pm\sqrt{m^2-c^4}$
is real. In Fig. \ref{fig:riemann}c the analytic continuation of the bulk spectrum
$E_k=\pm\sqrt{m^2+k^4}$ is depicted as a Riemann surface and compared with the Riemann 
surface of $E_k=\pm\sqrt{m^2+k^2}$ in Fig. \ref{fig:riemann}a.
It should be noted that this analytic continuation creates a non-Hermitian matrix
in Eq. (\ref{ham_2}):
\beq
\pmatrix{
m & k^2 \cr
k^2 & -m \cr
}\to
\pmatrix{
m & \pm ic^2 \cr
\pm ic^2 & -m \cr
}
.
\eeq
There is no conflict within our approach though, since we calculate the eigenfunctions before the analytic continuation,
where the matrix is real symmetric. Then the analytic continuation of the eigenfunctions and their inner product 
yields the correct result for the non-Hermitian matrix, as it can be directly checked for this simple case.
The analytic continuation $k\to e^{\pm i\pi/4}c$ of the eigenfunctions yields
\beq
\label{wf1}
\phi^+_n(e^{\pm i\pi/4}cr)
=J_n(e^{\pm i\pi/4}cr)+ iY_n(e^{\pm i\pi/4}cr)
=c_ne^{
\mp cr/\sqrt{2}}e^{icr/\sqrt{2}}\left[\frac{e^{\mp i\pi/8}}{\sqrt{cr}}+O\left(\frac{1}{cr}\right)\right]
\]
\[
\phi^-_n(e^{\pm i\pi/4}cr)
=J_n(e^{\pm i\pi/4}cr)- iY_n(e^{\pm i\pi/4}cr)
=c_ne^{
\pm cr/\sqrt{2}}e^{icr/\sqrt{2}}\left[\frac{e^{\mp i\pi/8}}{\sqrt{cr}}+O\left(\frac{1}{cr}\right)\right]
.
\eeq
These functions grow (decay) exponentially with $cr>0$, where
boundary conditions select the unique solution. For instance, 
$\phi^{+}(e^{-i\pi/4}cr)$ and $\phi^{-}(e^{i\pi/4}cr)$
are valid as an edge mode inside a disk with radius $r_0<\infty$, where the mode decays
exponentially from the edge of the disk toward its
center, and $\phi^{+}(e^{i\pi/4}cr)$ and $\phi^{-}(e^{-i\pi/4}cr)$ are 
valid for a circular hole of radius 
$r_0$ in an infinite disk. 

The two vector components $a_1$, $a_2$ of the eigenfunctions $\Psi_\pm(kr)$ in 
Eq. (\ref{ansatz2}) do not depend on the polar angle $\alpha$. Thus, 
the $S^2$ field is uniform in space:
\beq
{\vec s}
=\frac{1}{|a_1|^2+|a_2|^2}\pmatrix{
2|a_1||a_2|\cos(\varphi_2-\varphi_1) \cr
2|a_1||a_2|\sin(\varphi_2-\varphi_1) \cr
|a_1|^2-|a_2|^2 \cr
}
,
\eeq
which describes a fixed semicircle on the unit sphere, beginning with $a_2=0$ 
at the North Pole, hitting at $a_1=a_2$ the equator and ending with $a_1=0$ 
at the South Pole. For the bulk modes $a_{1,2}$ are real such that 
$\varphi_2-\varphi_1=0$ for $E\ge m$ and $\varphi_2-\varphi_1=\pi$ for $E\le -m$:
\beq
\label{2L_bulk}
{\vec s}_{\rm bulk}
=\frac{1}{a_1^2+a_2^2}\pmatrix{
2a_1a_2\cr
0 \cr
a_1^2-a_2^2 \cr
}
.
\eeq
According to Eq. (\ref{2b_laplacian}), on the other hand, for the edge modes
$a_1$ is imaginary and $a_2$ is positive. This gives for $k\to e^{\pm i\pi/4}c$
two branches of edge modes with the phases $\varphi_1=\pm\pi/2$, $\varphi_2=0$ and
\beq
\label{2L_edge}
{\vec s}_{\rm edge}
=\frac{1}{|a_1|^2+a_2^2}\pmatrix{
0\cr
\mp 2|a_1|a_2 \cr
|a_1|^2-a_2^2 \cr
}
, \ 
|a_1|=c^2
,\ 
a_2=m\pm\sqrt{m^2-c^4}
.
\eeq
Thus, the bulk and the edge modes are associated with semicircles on the unit sphere,
which meet the equator
at $(a_2/|a_2|,0,0)^T$ or at $(0,\pm 1,0)^T$, respectively, as illustrated by the red curves in 
Fig. \ref{fig:sphere}. In both cases 
the EV vanishes due to $\nabla\times{\vec s}=0$, while the divergence of ${\vec s}$ 
with respect to $\bk$ does not vanishes in the $1-2$ plane.
Finally, the intensity $(|a_1|^2+|a_2|^2|)|\phi^\pm|^2$ as a function of the radius $r$ decays like $1/r$
for the bulk modes but is strongly localized at the edge for the edge mode.

\subsection{Dirac model} 
\label{sect:dirac}

Another example is a Dirac Hamiltonian $H_D={\vec h}_D\cdot{\vec\sigma}$, 
which reads with polar coordinates
\beq
{\vec h}_D=(\cos\alpha\ i\partial_r-\sin\alpha\frac{1}{r}\ i\partial_\alpha,
\cos\alpha\frac{1}{r}\ i\partial_\alpha+\sin\alpha\ i\partial_r,m)
.
\eeq
It acts on a circular geometry and has the eigenfunctions (cf. App. \ref{app:Dirac_eigenvector})
\beq
\label{sols0}
\Psi_{k,n}(r,\alpha)=A_{k,n}\pmatrix{
C_n(kr) \cr
\rho(m,k) C_{n+1}(kr) e^{i\alpha}\cr
}
e^{in\alpha}
,
\eeq
where $C_n$ can be expressed as the linear combination of Bessel functions 
$\phi^\pm_n$ of Eq. (\ref{wf0}).
$\rho$, $E$ and $k$ are connected by two conditions (see Eq. (\ref{sign_relation1}) in 
App. \ref{app:Dirac_eigenvector}):
\beq
\label{sign_relation}
ki\rho=E-m
, \ \
k=i\rho(E+m)
,
\eeq
which implies $\rho^2=(m-E)/(m+E)$ and $k^2=E^2-m^2$.
The angular quantum numbers are $n=0,\pm 1,\ldots $, and the signs of $k$ and $\rho$ are 
fixed by the relations (\ref{sign_relation}). Thus, the eigenvalues are degenerate with respect to $n$,
since $H_D$ is circular invariant. In other words, $n$ is the eigenvalue of the angular momentum
operator, and the latter commutes with $H_D$.

\begin{figure}[t]
\begin{center}
bulk modes in the upper band:\\
\vskip0.2cm
\includegraphics[width=0.9\linewidth]{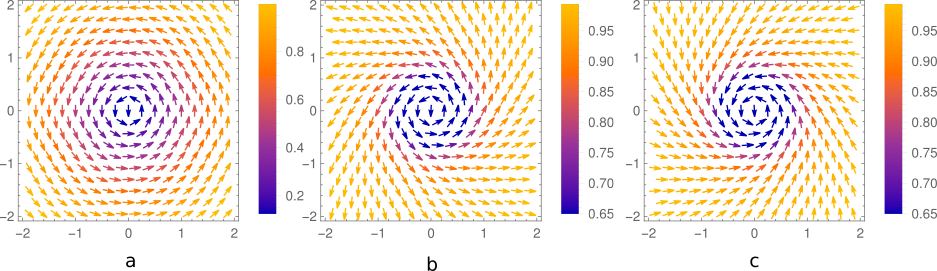}\\
edge modes:\\
\vskip0.3cm
\includegraphics[width=0.9\linewidth]{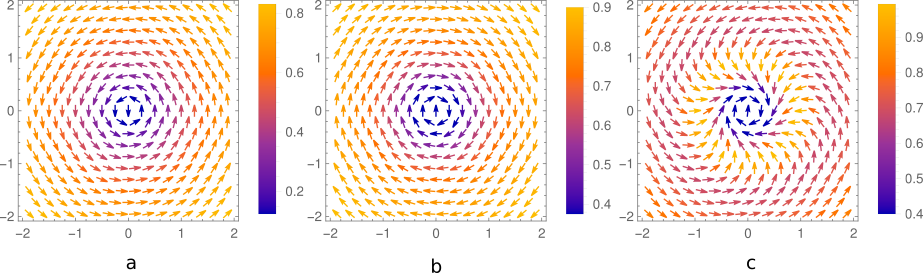}
\caption{
The $1-2$ projection of the $S^2$ field ${\vec s}$ of the Dirac model on a circular geometry.
Top row: the upper bulk modes for eigenfunctions based on a) $J_{1,2}$, b) $\phi^+_{1,2}$, 
and c) $\phi^-_{1,2}$.
Bottom row: the edge modes for eigenfunctions based on a) $J_{1,2}$, b) $\phi^+_{1,2}$, and c) $\phi^-_{1,2}$.
}
\label{fig:bulk_edge_mode}
\end{center}
\end{figure}

\begin{figure}[t]
\begin{center}
\includegraphics[width=0.9\linewidth]{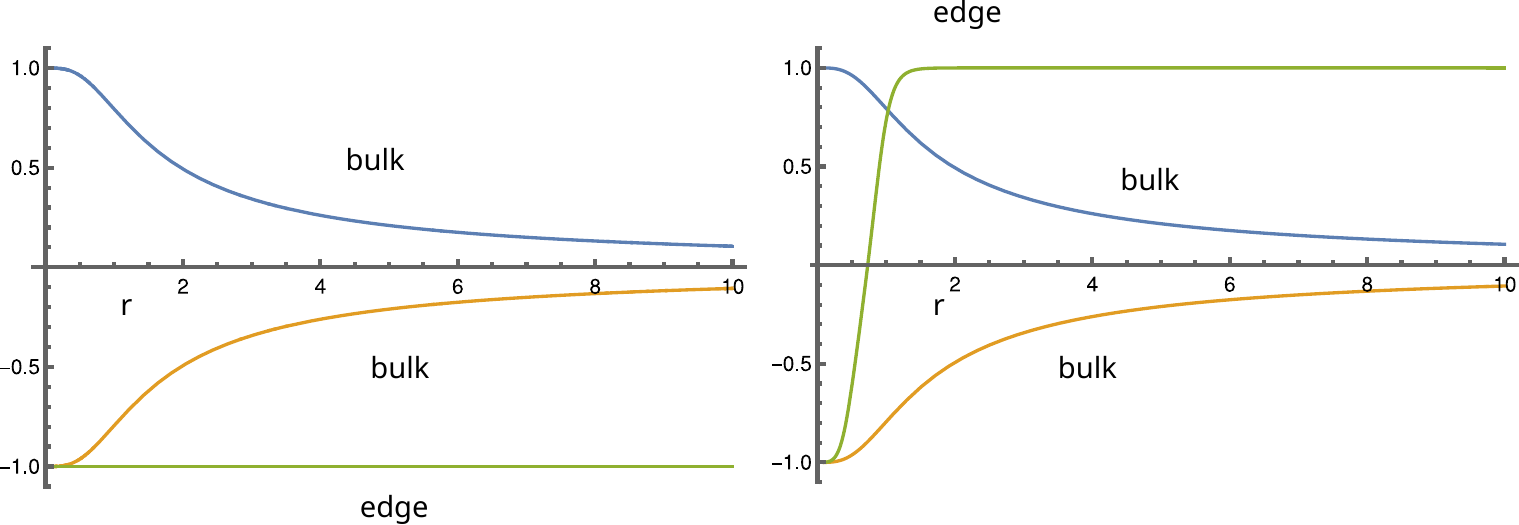}\\
\caption{
$\sin\eta_b$ as a function of the radius $r$ for the Dirac model with 
$\phi^+_{1,2}$ (left) and $\phi^-_{1,2}$ (right).
}
\label{fig:sin_eta}
\end{center}
\end{figure}

\begin{figure}[t]
\begin{center}
\includegraphics[width=1.0\linewidth]{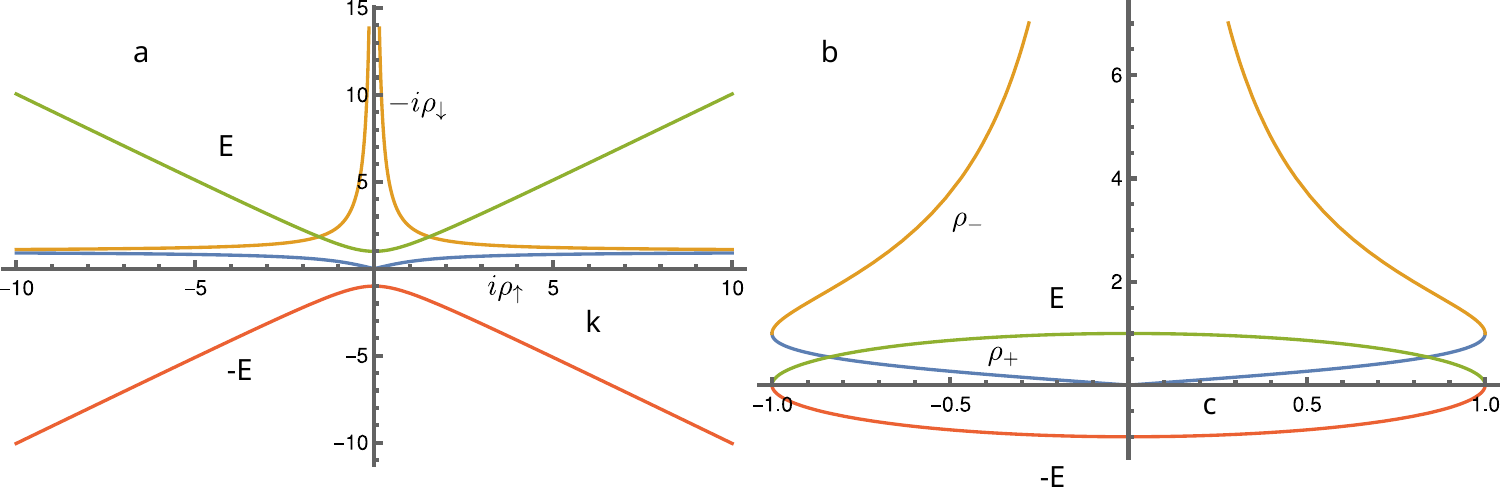}
\caption{
The parameter $\rho_b(k)$ and the eigenvalues $E$ for the bulk (a) 
and for the edge modes (b) of the Dirac model with $m=1$. $k=0$ and $c=0$ are the critical points $E=\pm 1$ 
for the localization-delocalization transition as well as as a significant change of the parameter $\rho_b$ 
(cf. Eqs. (\ref{rho_1}), (\ref{rho_2})).
}
\label{fig:rho}
\end{center}
\end{figure}

For $E>m$ and $E<-m$ $\rho$ is imaginary and $k=\pm\sqrt{E^2-m^2}$ is real. 
On the other hand, for $-m<E<m$ $\rho$ is real and $k=\pm\sqrt{E^2-m^2}$ is imaginary. 
In particular, for $E=0$ we have either $\rho=1$, $k=-im$ or $\rho=-1$, $k=im$.
This again is the analytic continuation $k\to ic$ for bulk to edge modes with 
$0\le c^2\le m^2$, illustrated in Fig. \ref{fig:riemann}b. Thus, from Eq. (\ref{wf0}) we obtain
edge modes with eigenvalues $E_c=\pm\sqrt{m^2-c^2}$, which either grow or decay exponentially with $r>0$
on the scale $1/c=1/\sqrt{m^2-E^2}$ for $-m< E< m$.
This scale diverges as $E$ approach the spectral boundaries $\pm m$ of the edge modes. 
On the other hand, the wavenumber of the bulk states vanishes as $k=\sqrt{E^2-m^2}$
for $E^2\sim m^2$.
Therefore, $E=\pm m$ are singular points in the spectrum, where the edge modes become uniformly
extended over the entire 2D system.
This is reminiscent of a localization-delocalization (or Anderson) transition in disordered systems 
and reflects the critical energy dependence of the edge modes and its transition to a bulk mode.
These results suggest the introduction of an index $b$ in the wavefunction of Eq. (\ref{sols0}) 
\beq
\label{sols1}
\Psi_{k,n,b}(r,\alpha)=A_{k,n,b}\pmatrix{
C_n(kr) e^{-i\alpha/2}\cr
\rho_b(m,k)C_{n+1}(kr) e^{i\alpha/2}\cr
}e^{i(n+1/2)\alpha}
,
\eeq
where $b$ is either the band index $b=\u,\d$ for the upper and the lower band,
or $b=\pm$ for the energies $E=\pm\sqrt{m^2-c^2}$ of the edge modes. 
This gives for the $S^2$ field
\beq
\label{s_field1}
{\vec s}_b=\frac{1}{|C_n|^2+|\rho_b C_{n+1}|^2}\pmatrix{
2|C_n^*C_{n+1}\rho_b|\cos(\alpha+\eta_b)\cr
2|C_n^* C_{n+1}\rho_b|\sin(\alpha+\eta_b)\cr
|C_n|^2-|C_{n+1}\rho_b|^2 \cr
}
\ \ {\rm with}\ \
\eta_b={\rm arg}(C_n^*C_{n+1}\rho_b)
.
\eeq
The $1-2$ projection is visualized for several eigenfunctions in Fig. \ref{fig:bulk_edge_mode}.
The phase shift $\eta_b$ is the angle between the radial vector $\br$ and
the projected vector $(s_1,s_2)^T$.
Comparing this expression with Eq. (\ref{s3-field}), we get
the phase relation between the eigenfunctions and the $S^2$ field as
\beq
\varphi_2-\varphi_1=\alpha+\eta_b
.
\eeq
Besides the different energies for the bands of the bulk and for the edge modes,
also the parameter $\rho_b(m,k)$ distinguishes the different modes. 
With the band energies $E=\pm\sqrt{m^2+k^2}$ we get from the 
relations in Eq. (\ref{sign_relation}) for the bulk an imaginary parameter
\beq
\label{rho_1}
\rho_\u=-i\sqrt{\frac{\sqrt{m^2+k^2}-m}{\sqrt{m^2+k^2}+m}}
=1/\rho_\d
.
\eeq
On the other hand, for the edge modes we have a real parameter
\beq
\label{rho_2}
\rho_+=\sqrt{\frac{m-\sqrt{m^2-c^2}}{m+\sqrt{m^2-c^2}}}
=1/\rho_-
.
\eeq
This entails that under $m\to-m$ we have $\rho_\u\leftrightarrow -\rho_\d$
and $\rho_+\leftrightarrow\rho_-$.
The behavior of the $\rho_b(k)$ is presented in Fig. \ref{fig:rho}.
Since ${\rm arg}(\rho_b)$ is independent of the values of $m$ as well as $k$ and $c$,
respectively, it can be understood as a topological number, analogous to
the Chern number, which is associated with the band:
\beq
\label{top_index2}
{\rm arg}(\rho_b)=\cases{
-\pi/2 & $b=\u$ \cr
\pi/2 & $b=\d$ \cr
0 & $b=\pm$ \cr
}
.
\eeq
The phase shift $\eta_b$ in ${\vec s}_b$ of Eq. (\ref{s_field1}) has different 
values for the two bands and for the edge modes due to 
$\eta_b={\rm arg}(C_n^*C_{n+1})+{\rm arg}(\rho_b)$. $\sin\eta_b(r)$ is plotted 
in Fig. \ref{fig:sin_eta} for bulk and edge modes and for different eigenfunctions.

There are transitions from bulk to edge modes at $E=m$ and from edge modes to bulk modes
at $E=-m$. They are associated with a change
of ${\rm arg}(\rho_b)$ or by a sign change of $\sin\eta_b$. The latter appears in the EV as
\beq
{\cal V}
=\int_0^{2\pi} {\vec s}\cdot \frac{d\br}{d\alpha}d\alpha
=4\pi r_0\frac{|C_n^*C_{n+1}\rho_b|}{|C_n|^2+|\rho_bC_{n+1}|^2} \sin\eta_b(r_0)
\eeq
after an integration with respect to the edge of a circular hole with radius $r_0$.
EV is positive in the upper band and negative in the lower band (cf. Fig. \ref{fig:sin_eta}).
Since $\eta_b$ depends on the radius, the EV can change with the position of the edge
and the boundary conditions.

\begin{figure}[t]
\begin{center}
\includegraphics[width=0.9\linewidth]{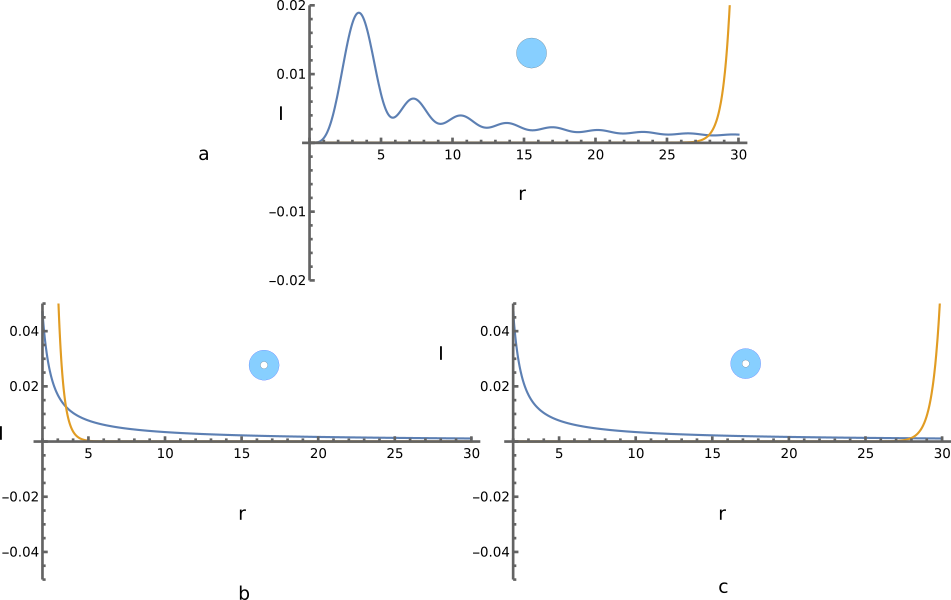}
\caption{
Radial intensity distribution $I(r)$ for the Dirac model on a disk with radius $r_0=30$
for bulk (blue) and edge modes (yellow). The eigenfunctions are based on
a) $J_{1,2}$, b) $\phi^+_{1,2}$, and c) $\phi^-_{1,2}$.
}
\label{fig:intens}
\end{center}
\end{figure}

$\rho_\d$ and $\rho_-$ diverge for $k\to0$, while $\rho_\u$ and $\rho_+$ vanish in this limit
(cf. Fig. \ref{fig:rho}), reflecting the qualitative change of the wavefunctions from edge-localized 
functions to circular waves at $E=\pm m$. At the critical point $E=-m$
the $S^2$ vector reaches the South Pole. On the other hand,
at the critical point $E=m$ the $S^2$ vector arrives at the North Pole.
The trajectories of ${\vec s}(\alpha)$ at fixed $r$ are indicated by the blue circles in Fig. \ref{fig:sphere}.
Finally, the intensity $I=|C_n|^2+|\rho_b C_{n+1}|^2$ is plotted in Fig. \ref{fig:intens}
for different eigenfunctions.
In all three cases the main difference is the significant weight of the edge mode intensity
either at the center of the disk or at its boundary.


\section{Discussion and conclusions}
\label{discussion}

After having solved the eigenvalue equation for the bulk modes, 
the analytic continuation of the real wavenumber $k$ into the complex plane 
provides the edge modes. This approach yields a spectral separation of bulk and 
edge modes. On the other hand, there
is no complete spatial separation of these two types of modes, since the edge modes
extend into the bulk with an exponential decay on the scale $\xi$:
\beq
\xi=\cases{
1/\sqrt{|E|} & single Laplacian \cr
\sqrt{2}/|(m^2-E^2)^{1/4}| & two-band Laplacian model \cr
1/\sqrt{m^2-E^2} & Dirac model \cr
}
.
\eeq
This means that the edge modes are two-dimensional, which cannot be described
as solutions of a one-dimensional equation. However, they do not spread over the entire
system but decay exponentially from the edge. In contrast, the bulk modes decay like $1/\sqrt{r}$.

By varying $E$ we can scan continuously through the bulk and the edge spectrum.
There are bulk-edge transitions at $E=\pm m$, representing transitions between localized edge modes
and delocalized bulk modes.
They are accompanied by a qualitative change for the Dirac model, since the index ${\rm arg}(\rho_b)$ 
and the sign of $\sin\eta_b$ change according to Eq. (\ref{top_index2}) and Fig. 
\ref{fig:sin_eta}.
For the two-band Laplacian model the phase difference $\Delta\varphi:=\varphi_2-\varphi_1$ switches
from $\Delta\varphi=0$ for the upper band and $\Delta\varphi=\pi$ for the lower band 
to $\Delta\varphi=\pm \pi/2$ for the two branches of the edge modes. 
This indicates that the bulk and the edge modes are not only characterized by their
spatial decay but also by the phase difference $\Delta\varphi$ in the $S^2$ field ${\vec s}$.

\subsection{Boundary conditions}
\label{sect:bc}

Next, we introduce boundary conditions for the eigenfunctions of the Hamiltonian, whose purpose is two-fold, 
namely (i) to get unique solutions for the physical system (i.e., there can be no ambiguity in terms of physical 
results) and (ii) to obtain self-adjoint operators for a given Hilbert space:
Although we have found the solution of $H\Psi_E=E\Psi_E$, to diagonalize the Hamiltonian matrix
we still need that the inner product $(\Psi_E,\Psi_{E'})=0$ for $E\ne E'$. This is equivalent with the request
that $H$ is self-adjoint. For instance, the Laplacian $\Delta$ is diagonal if
the boundary terms in
\beq
(\Psi_E,\Delta\Psi_{E'})=(\Delta\Psi_E,\Psi_{E'})+{\rm boundary\ terms}
\eeq
vanish due to appropriate boundary conditions.
We can also directly obtain such a relation by using the fact that we operate in the eigenbasis of the
Laplacian for a disk geometry with radius $R$, using Bessel functions. Thus, the Hilbert space is spanned by 
Bessel functions on the interval $0\le r\le R$, where the inner product reads 
\beq
(\psi_{kn},\psi_{qn'})=
\int_0^{2\pi}e^{i(n'-n)\alpha}d\alpha\int_0^R\phi_{n}(kr)\phi_{n'}(qr)rdr
=\delta_{nn'}\int_0^R\phi_{n}(kr)\phi_{n}(qr)rdr
\eeq
with $\phi_n=J_n+AY_n$ and a real coefficient $A$.
For a self-adjoint Hamiltonian this relation requires the orthogonality of the eigenfunctions; i.e., $\delta_{kq}$
for the integral on the right-hand side. 
The latter can be obtained with proper boundary conditions (rule 11.4.3 in Ref.~\cite{abramowitz+stegun})
since
\beq
\int_{0}^{R}\phi_{n}(kr)\phi_{n}(qr)rdr=\frac{R}{q^2-k^2}
\left[k\phi_n(qR)\phi'_n(kR)-q\phi_n(kR)\phi'_n(qR)\right]
.
\eeq
Then the Bessel functions are orthogonal, for instance, for the zeros of $\phi_n$ or of $\phi'_n$.
In other words, there is a sequence $\{k_{n,l}\}_l$ of wave numbers with $\phi_n(k_{n,l}R)=0$
that provides the orthogonal eigenfunctions of the Laplacian. Another sequence $\{k'_{n,l}\}$
is obtained from $\phi'_n(k'_{n,l}R)=0$ as boundary condition. Corresponding boundary conditions
for the Dirac operator are obtained for $\{k_{nl}\}$, which are solutions of
\beq
A=-\frac{J_n(k_{nl}R)}{Y_n(k_{nl}R)}
=-\frac{J_{n+1}(k_{nl}R)}{Y_{n+1}(k_{nl}R)}
\eeq 
and imply $\phi_n(k_{nl}R)=\phi_{n+1}(k_{nl}R)=0$.

In the next step we apply boundary conditions to the edge modes on a disk with radius $R$. 
Starting from general bulk eigenfunctions of the Laplacian, we perform the analytic continuation 
of $k\to\pm ic$. Assuming bounded solutions, we are enforced to use $I_n(c r)$ because $K_n(c r)$ 
is singular at $r=0$. Since $I_n(c r)$ is a monotonic increasing function, only mixed 
boundary conditions  can be satisfied: $I_n(c R)-bc I'_n(c R)=0$ with a real parameter $0<b<\infty$, 
which can also be written with $z=c R$ as
\beq
\frac{d\log I_n(z)}{dz}=\frac{I'_n(z)}{I_n(z)}=\frac{1}{bc}=\frac{R}{bz}
.
\eeq
Since $d\log(I_n(z))/dz$ increases monotonically, while the right-hand side decreases monotonically,
there is only a single solution $c_n$ for this boundary condition at fixed $n$. 
Thus, after fixing the boundary conditions we have a point spectrum $\pm\sqrt{m^2-c_n^2}$ on 
the circle in Fig. \ref{fig:riemann}b, and  there is no continuous transition from the edge to the bulk 
spectrum.

\subsection{Properties of the $S^2$ field }

The $S^2$ real vector field ${\vec s}$ characterizes the gauge-invariant properties of the eigenfunctions.
We briefly summarize its behavior, visualized by the unit sphere in Fig. \ref{fig:sphere}. Under the variation
of the polar coordinates $(r,\alpha)$ the trajectory of the unit vector depends strongly on the model.
For the two-band Laplace model there is no $\alpha$ dependence, such that the trajectory follows the
meridians. The meridian trajectory jumps by an angle of $\pi/2$ when we switch from bulk to edge modes, as
stated in Eqs. (\ref{2L_bulk}) and (\ref{2L_edge}). This reflects the fact that the model is topological
trivial. In contrast, the trajectories of Dirac Hamiltonian in Eq. (\ref{s_field1}) cover the entire sphere, where a 
variation of $r$
follows also the meridians, while a variation of $\alpha$ for $0\le\alpha<2\pi$ gives a full circle at fixed
latitude, reflecting that the winding number of ${\vec s}$ is 1.  Going from bulk to edge states at fixed $\alpha$
results in a jump of the meridian, similar to the two-band Laplace model. Thus, the winding number of ${\vec s}$
can be used for the characterization of the Hamiltonian. This should also be applicable to Hamiltonians
with more than two bands.

\subsection{Comparison with the TBEC approach}

The TBEC for a straight (infinite) edge was studied 
for the 2D Dirac model \cite{2403.04465} and for the extended hydrodynamic model  \cite{2410.13940},
using the relation between the Chern number $C_+$ of the bulk modes in the upper band.
The TBEC holds if the number $n_e$ of edge modes 
is equal to $C_+$. For the translation-invariant 2D Dirac operator the Chern number reads
\beq
C_+=\frac{1}{2\pi i}\int_{\mathbb{R}^2} {\rm tr}(P_+[\partial_{k_x}P_+,\partial_{k_y}P_+])dk_xdk_y
=1
,
\eeq
where $P_+$ is the eigenprojection for the upper band. 
Depending on the boundary conditions, though, the TBEC does not always hold, since $n_e=2,3$
for some boundary conditions~\cite{2403.04465}. The origin of this violation of the TBEC has been associated 
with the unbounded 
spectrum of the translation-invariant 2D Dirac operator. In contrast, the edge is finite for the circular-invariant 
2D Dirac operator, implying that also its spectrum is bounded. The difference between the straight edge
and the circular edge can be seen already in the spectrum before employing boundary conditions: 
for an edge along the $x$-axis we have
$E(k_x,c)=\pm\sqrt{m^2+k_x^2-c^2}$ with $-\infty< k_x<\infty$, $c^2\le m^2+k_x^2$,
while $E(n,c)=\pm\sqrt{m^2-c^2}$ with $c^2\le m^2$ for the circular edge is bounded.
Another example for the violation of the TBEC appears for the two-band Laplacian, where $C_+=0$. 
As discussed in Sect. \ref{sect:2_l} and in Sect. \ref{sect:bc}, there are edge modes.  It should be noted that
the ABEC does not lead to any conflict in this case. However, the counting of the number of edge modes 
that merge with the bulk modes through the Chern number is not possible here.

\subsection{Robustness}

An important question concerns the robustness of the edge modes when we replace the uniform 
mass $m$ by a spatially varying mass $m(r,\alpha)$. In the circular-symmetric case $m(r)$,
a sign change of the mass creates an additional edge mode with a skyrmion-like wavefunction~\cite{Ziegler:18}.
Moreover, with a positive $m(r,\alpha)={\bar m}+\delta m(r,\alpha)$ we can break the circular symmetry.
If $\delta m(r,\alpha)/{\bar m}\ll 1$, the robustness of the edge modes can be analyzed within perturbation 
theory. The (degenerate) perturbation expansion in powers of $\delta m(r,\alpha)$ would provide a stability
analysis of the edge modes. A thorough study of this perturbation approach exceeds the scope of
this paper and should be left for a separate project in the future.

\subsection{Conclusions}

As a summary of the results of the two-band examples, we found the bulk eigenfunctions 
of circular symmetric two-band models. They are of the form 
\[
\Psi_{k,n}(r,\alpha)
=\pmatrix{
\phi_1 \cr
\phi_2 \cr
}e^{ikr+in\alpha}
\]
with $k$ real and $\phi_{1,2}$ complex. Physical properties are obtained from the $S^2$ field ${\vec s}$ 
\[
\frac{\Psi_{k,n}\cdot{\vec\sigma}\Psi_{k,n}}{\Psi_{k,n}\cdot\Psi_{k,n}}
=\frac{|\phi_1^*\phi_2|}{|\phi_1|^2+|\phi_2|^2}\pmatrix{
\cos(\Delta\varphi) \cr
\sin(\Delta\varphi) \cr
|\phi_1|^2-|\phi_2|^2 \cr
}
\ ,\ \ \Delta\varphi={\rm arg}(\phi_1^*\phi_2)
.
\]
The $1-2$ projection of the $S^2$ field identifies the phase difference of the 
two components of the eigenfunctions. An analytic continuation of $k$ yields
the corresponding expressions for the edge modes.
From these results we conclude that the ABEC offers a systematic approach 
through an analytic continuation of the wavenumber for the description of 
two-band Hamiltonians with edges. Although we have focused here on a circular 
geometry for simplicity, the concept can be extended to other geometries 
with finite edges.

\vskip0.3cm

\no
{\bf Acknowledgment}

\no
I am grateful to the anonymous Referee for bringing the works of Refs. \cite{hatsugai93,2403.04465,2410.13940} 
to my attention.

\appendix

\section{Eigenfunctions of the Dirac Hamiltonian}
\label{app:Dirac_eigenvector}

With the ansatz
\beq
\label{ansatz}
\Phi_{k,n}(r,\alpha)=\pmatrix{
f_n(kr) \cr
g_n(kr) e^{i\alpha}\cr
}
e^{in\alpha}
\eeq
we can write for the eigenvalue problem $H_D\Phi_{E,n}=E\Phi_{E,n}$ the equation
\beq
\pmatrix{
0 & ie^{-i\alpha}(\partial_r-\frac{i}{r}\partial_\alpha) \cr
ie^{i\alpha}(\partial_r+\frac{i}{r}\partial_\alpha) & 0 \cr
}\pmatrix{
f_n(kr) \cr
g_n(kr) e^{i\alpha}\cr
}
e^{in\alpha}
\]
\[
=\pmatrix{
i(kg_n\rq{}+\frac{n+1}{r}g_n)e^{in\alpha} \cr
i(kf_n\rq{}-\frac{n}{r}f_n)e^{i(n+1)\alpha} \cr
}
=\pmatrix{
(E-m)f_n e^{in\alpha}\cr
(E+m)g_n e^{i(n+1)\alpha}\cr
}
,
\eeq
which enables us to eliminate $e^{in\alpha}$ and $e^{i(n+1)\alpha}$, respectively, on both sides
of the second equation:
\beq
\label{eig_eq2}
\pmatrix{
i(kg_n\rq{}+\frac{n+1}{r}g_n) \cr
i(kf_n\rq{}-\frac{n}{r}f_n) \cr
}
=\pmatrix{
(E-m)f_n\cr
(E+m)g_n \cr
}
.
\eeq
Now we consider that $C_n$ is either the Bessel function $J_n$, $Y_n$ or a linear combination
of these two functions and
write $f_n(kr)=C_n(kr)$ and $g_n(kr)=\rho C_{n+1}(kr)$.
Then we employ the recurrence relations of the Bessel functions~\cite{abramowitz+stegun}
\beq
C_n'(r)=-\frac{n}{r}C_n(r)+C_{n-1}(r)
\ \ (n=1,2,...)
\eeq
and
\beq
C_n'(r)=\frac{n}{r}C_n(r)-C_{n+1}(r)
\ \ (n=0,1,...)
.
\eeq
With these relations we obtain from Eq. (\ref{eig_eq2})
\beq
\pmatrix{
ik\rho C_n \cr
-ikC_{n+1} \cr
}
=\pmatrix{
(E-m)C_n\cr
(E+m)\rho C_{n+1}\cr
}
\eeq
or the eigenvalue equation
\beq
\pmatrix{
m+ik\rho & 0 \cr
0 & -m-ik/\rho \cr
}\pmatrix{
C_n \cr
\rho C_{n+1} \cr
}=E\pmatrix{
C_n \cr
\rho C_{n+1} \cr
}
,
\eeq
which gives the relations
\beq
\label{sign_relation1}
k\rho=i(m-E)
, \ \
k=i(E+m)\rho
.
\eeq
This determines the parameter $\rho$ in $g_n$.
Thus, we have $k^2=E^2-m^2$ and $\rho^2=(m-E)/(m+E)$, both are independent of $n$.
The eigenfunction in Eq. (\ref{ansatz}) becomes
\beq
\Phi_{k,n}(r,\alpha)=\pmatrix{
C_n \cr
\rho C_{n+1}e^{i\alpha} \cr
}e^{in\alpha}
.
\eeq

\section{Analytic continuation of Bessel functions}
\label{app:bessel}

The analytic continuation $k\to ic$ with a real $c$ yields for
the Bessel functions~\cite{abramowitz+stegun}
\beq
J_n(kr)\to J_n(icr)=e^{i\pi n/2}I_n(cr)
\ ,\ \
H_n^{(1)}(kr)
\to H_n^{(1)}(icr) 
=-\frac{2i}{\pi}e^{-i\pi n/2}K_n(cr)
\eeq
with $H_n^{(1)}(kr)=J_n(kr)+iY_n(kr)$,
which implies
\beq
\label{bj}
J_n(icr)^*J_{n+1}(icr)
=iI_n(cr)I_{n+1}(cr)
\eeq
and
\beq
\label{hankel}
H_n^{(1)}(icr)^*H_{n+1}^{(1)}(icr)
=-\frac{4i}{\pi^2}K_n(cr)K_{n+1}(cr)
.
\eeq

\end{document}